\documentclass[aps,pre,twocolumn,titlepage]{revtex4}

\usepackage{graphicx}% Include figure files
\usepackage{color}
\usepackage{dcolumn}% Align table columns on decimal point
\usepackage{latexsym}
\usepackage[normalem]{ulem}
\usepackage{hyperref,amssymb}
\usepackage{url}
\usepackage{color,soul}
\usepackage{graphicx}
\usepackage{verbatim}
\usepackage{multirow}
\usepackage{amsmath}
\newcommand{\beq}{\begin{eqnarray}}
\newcommand{\eeq}{\end{eqnarray}}
\usepackage{mathrsfs}
\usepackage{float,soul}
\usepackage[dvipsnames]{xcolor}
\usepackage{mathtools}
\usepackage{slashed}
\usepackage{physics}	% installed packages for typing maths and physics
\usepackage{graphicx}   % need for Fig.ures
\usepackage{epstopdf}
\usepackage{subfigure}  % use for side-by-side Fig.ures
\usepackage{hyperref}   % use for hypertext links, including those to external documents and URLs
\usepackage{bbold}
\usepackage{wasysym}
\usepackage{feynmp}
\usepackage{hyperref}
\newcommand{\angstrom}{\textup{\AA}}
\usepackage{lipsum}
\usepackage[normalem]{ulem}
\usepackage{hyperref,amssymb}
\usepackage{url}
\usepackage{color,soul}
\usepackage{graphicx}
\usepackage{verbatim}
\usepackage{multirow}
\usepackage{amsmath}
\usepackage{mathrsfs}
\usepackage{float,soul}
\usepackage{mathtools}
\usepackage{slashed}
\hypersetup{colorlinks,
}

\usepackage[latin1]{inputenc}

\begin{document}

\title{\Large Shaping the topology of twisted bilayer graphene\\ via time-reversal symmetry breaking}
\author{Cunyuan Jiang$^{1,2,3}$}\author{Matteo Baggioli$^{1,2,3}$}
\email{b.matteo@sjtu.edu.cn}
\author{Qing-Dong Jiang$^{1,4,5}$}
\email{qingdong.jiang@sjtu.edu.cn}
\address{$^1$ School of Physics and Astronomy, Shanghai Jiao Tong University, Shanghai 200240, China}
\address{$^2$ Wilczek Quantum Center, School of Physics and Astronomy, Shanghai Jiao Tong University, Shanghai 200240, China}
\address{$^3$ Shanghai Research Center for Quantum Sciences, Shanghai 201315,China}
\address{$^4$ Tsung-Dao Lee Institute, Shanghai Jiao Tong University, Shanghai 200240, China}
\address{$^5$ Shanghai Branch, Hefei National Laboratory, Shanghai 201315, China}

\begin{abstract}
Symmetry breaking is an effective tool for tuning the transport and topological properties of 2D layered materials. Among these materials, twisted bilayer graphene (TBG) has emerged as a promising platform for new physics, 
characterized by a rich interplay between topological features and strongly correlated electronic behavior. In this study, we utilize time-reversal symmetry breaking (TRSB) to manipulate the topological properties of TBG. By varying the strength of TRSB, we discover a topological phase transition between a topological insulating phase, which exhibits a pair of flat bands with opposite Chern numbers, and a novel insulating state where the Chern number, but not the Berry curvature, of the flat bands vanishes. We demonstrate that this topological transition is mediated by a gap closing at the $\Gamma$ point, and we construct a three-dimensional phase diagram as a function of the twisting angle, the symmetry-breaking parameter, and the mismatch coupling between AA and AB stacking regions. Finally, we show that this novel electronic phase can be identified in the lab by measuring, as a function of the Fermi energy, its non-quantized anomalous Hall conductivity that is induced by the Berry dipole density of the lowest flat bands.
\end{abstract}

\maketitle
Twisted bilayer graphene (TBG) is a 2D quantum material composed of two stacked layers of graphene twisted with a relative angle $\theta$ that create a so-called moir\'e pattern \cite{Andrei2020}. TBG has attracted a lot of attention in the recent years because of strongly-correlated electronic phases and anomalous properties emerging around the so-called ``magic-angle'', $\theta \approx 1.09^\circ$ \cite{gibney2019magic}. Interestingly, TBG hosts a pair of topological electronic flat bands and provides a perfect platform to explore the interplay between topology, strongly correlated physics and exotic superconductivity.

On the other hand, symmetry breaking represents a valuable tool for controlling and engineering the physical properties of 2D materials, including their topological structure \cite{Du2021}. TBG is no exception. Time-reversal symmetry breaking (TRSB) is particularly important because it enables the emergence of several physical phenomena that are otherwise forbidden. In particular, the Berry curvature $\Omega(\boldsymbol{q})$ (with $\boldsymbol{q}$ being the wave vector) flips its sign under time-reversal operation, \textit{i.e.}, $\Omega(\boldsymbol{q}) = -\Omega(-\boldsymbol{q})$. Consequently, in a time-reversal symmetric system, the overall Berry curvature must vanish, leading to a zero anomalous Hall conductivity. 

TRSB in TBG can be introduced in various ways such as imposing an external magnetic field \cite{sanchez2024correlated}, exploiting coherent light-matter interactions \cite{PhysRevB.101.241408,PhysRevResearch.1.023031}, and using chiral optical cavities \cite{jiang2023engineering} or hBN encapsulation \cite{Long2022}. As for the simpler case of monolayer graphene \cite{PhysRevB.99.235156}, TRSB can induce topological phase transitions in TBG \cite{PhysRevResearch.1.023031} and can have strong effects on the electronic spectrum flattening for example the topological isolated bands away from the magic angle \cite{jiang2023engineering}. 

In this study, leveraging on an effective model of TBG, we study in detail the effects of TRSB in twisted bilayer graphene as a function of the twisting angle and the different interlayer coupling strength of AA and AB stacked regions \cite{PhysRevX.8.031087}, that is known to appear because of the surface relaxation effects and the consequent atomic corrugation \cite{PhysRevB.90.155451,PhysRevB.99.195419}. In doing so, we reveal a previously overlooked topological phase transition (TPT) that appears at a critical value of TRSB, and that leads to a new electronic state exhibiting a pair of flat bands with zero Chern number but non-trivial Berry curvature.

\begin{figure*}[ht!]
    \centering
    \includegraphics[width=\linewidth]{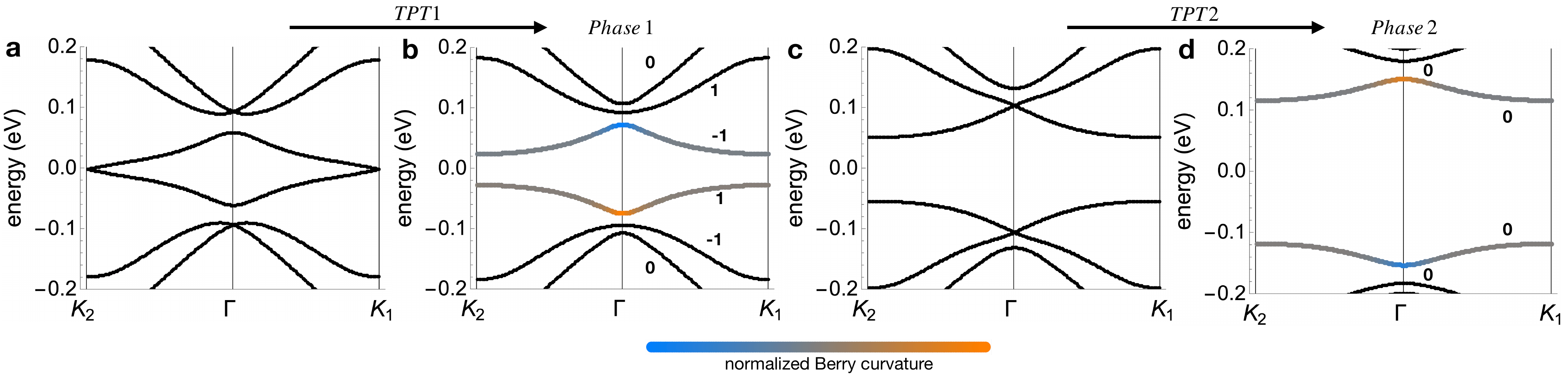}
    \caption{The electronic band structure of TBG for $u=0.817$, $u'=1$, $\theta=1.5^\circ$ by varying the strength of time-reversal symmetry breaking $\tau$ from $0$ \textbf{(a)} to $0.2$ \textbf{(d)}. The color map indicates the value of the normalized Berry curvature for the lowest bands that is shown more clearly in panel b of Fig.\ref{fig2}. The integers show the Chern numbers of the lowest electronic bands. }
    \label{fig1}
\end{figure*}

We model TBG using an effective Hamiltonian \cite{handbook} $H_{\text{TBG}}(\mathbf{q})$,

\begin{equation}
    \left(\begin{matrix}
    H_1 (\mathbf{q}) & T_{\mathbf{q}_b} & T_{\mathbf{q}_{tr}} & T_{\mathbf{q}_{tl}} & \cdots\\ T_{\mathbf{q}_b}^\dagger &  H_2 (\mathbf{q}-\mathbf{q}_b) & 0 & 0 & \cdots\\ T_{\mathbf{q}_{tr}}^\dagger & 0 & H_2 (\mathbf{q}-\mathbf{q}_{tr}) & 0  & \cdots\\ T_{\mathbf{q}_{tl}}^\dagger & 0 & 0 & H_2 (\mathbf{q}-\mathbf{q}_{tl}) \\ \vdots & \vdots & \vdots & \vdots & \ddots
    \end{matrix}\right),\label{h}
\end{equation}
where $\mathbf{q}$ is the wave-vector, and
\begin{equation}
    H_{n}(\mathbf{q})=v_F \left(\begin{matrix}
    0 & q_x-i q_y \\ q_x+i q_y & 0
    \end{matrix}\right)
\end{equation}
is the Hamiltonian of each decoupled layer. $n=1,2$ is the index of the layer and the Fermi energy is fixed to $v_F\approx 5.944 eV \cdot \angstrom$. We refer to Ref.~\cite{handbook} for more details about the precise definitions of the wave-vectors $\mathbf{q}_b,\mathbf{q}_{tr},\mathbf{q}_{tl}$.

Finally, the hopping matrix elements are given by
\begin{align}
   & T_{\mathbf{q}_{b}}= t\left(\begin{matrix}
    u & u' \\ u' & u
    \end{matrix}\right),\quad T_{\mathbf{q}_{tr}}= t\left(\begin{matrix}
    u e^{i \phi} & u' \\ u' e^{-i \phi} & u e^{i \phi}
    \end{matrix}\right),\nonumber \\
    &\hspace{1.5cm}T_{\mathbf{q}_{tl}}= t\left(\begin{matrix}
    u e^{-i \phi} & u' \\ u' e^{i \phi} & u e^{-i \phi}
    \end{matrix}\right),
\end{align}
with $\phi=2\pi/3$ and $t=0.11$. In order to take into account the different interlayer coupling strength of AA and AB stacked regions \cite{PhysRevX.8.031087}, we  fix $u'=1$ and keep $u \neq 1$ as a free parameter.

We then deform our initial system by adding a time-reversal breaking deformation
\begin{equation}
    H_{\text{TBG}+\tau}(\mathbf{q})=H_{\text{TBG}}(\mathbf{q})
    +\tau \sigma_z \mathbb{I},
    \label{h+t}
\end{equation}
where the parameter $\tau$ serves to quantify the strength of TRSB, $\sigma_z$ is the Pauli spin matrix and $\mathbb{I}$ the identity matrix in the multi-dimensional wave-vector space on which $H_{\text{TBG}}$ is defined.

We obtain the electronic band structure by direct diagonalization of $H_{\text{TBG}+\tau}$ keeping as free parameters the twisting angle $\theta$, the time-reversal breaking strength $\tau$ and the coupling strength $u$. Moreover, we compute the Berry curvature of the $n$th band at wave-vector $\boldsymbol{q}$ using the following expression 
\begin{equation}\label{dede}
    \Omega_n (\boldsymbol{q})=i[\langle\partial_{q_x} n(\boldsymbol{q})|\partial_{q_y} n(\boldsymbol{q})\rangle-\langle\partial_{q_y},n(\boldsymbol{q})|\partial_{q_x} n(\boldsymbol{q})\rangle],
\end{equation}
where $|n(\boldsymbol{q})\rangle$ labels the eigenstate of $n$th band at wave vector $\boldsymbol{q}$. Finally, we define the topological Chern number $\mathcal{C}$,
\begin{equation}
    \mathcal{C}_n=\frac{1}{2\pi} \int_{\text{BZ}} \Omega_n (\boldsymbol{q}) d\boldsymbol{q},
\end{equation}
where the integration is taken over the whole Brillouin zone (BZ).

We start by exploring the effects of TRSB at a constant twisting angle $\theta=1.5^\circ$ and coupling $u=0.817$. In Fig.\ref{fig1}, we show the electronic band structure of TBG by dialing the symmetry breaking parameter $\tau$. Panel (a) corresponds to the time-reversal symmetric case ($\tau=0$), where the band structure displays a pair of weakly dispersing electronic bands crossing at $K$ points. In this case, both the Berry curvature and the Chern number of all the bands identically vanish. Therefore, $\tau=0$ corresponds to a topologically trivial phase.

By breaking time-reversal symmetry, $\tau \neq 0$ (panel (b) in Fig.\ref{fig1}), a gap opens at the $K$ points and two isolated bands emerge in the spectrum with non-trivial and opposite topological Chern number $\mathcal{C}=\pm 1$. The system is now in a topological insulating phase (phase 1). This first topological phase transition (TPT1 in Fig.\ref{fig1}) is not new \cite{PhysRevResearch.1.023031,jiang2023engineering}, and it is indeed analogous to the case of monolayer graphene under time-reversal symmetry breaking \cite{PhysRevB.99.235156}.

\begin{figure*}[htb]
    \centering
    \includegraphics[width=\linewidth]{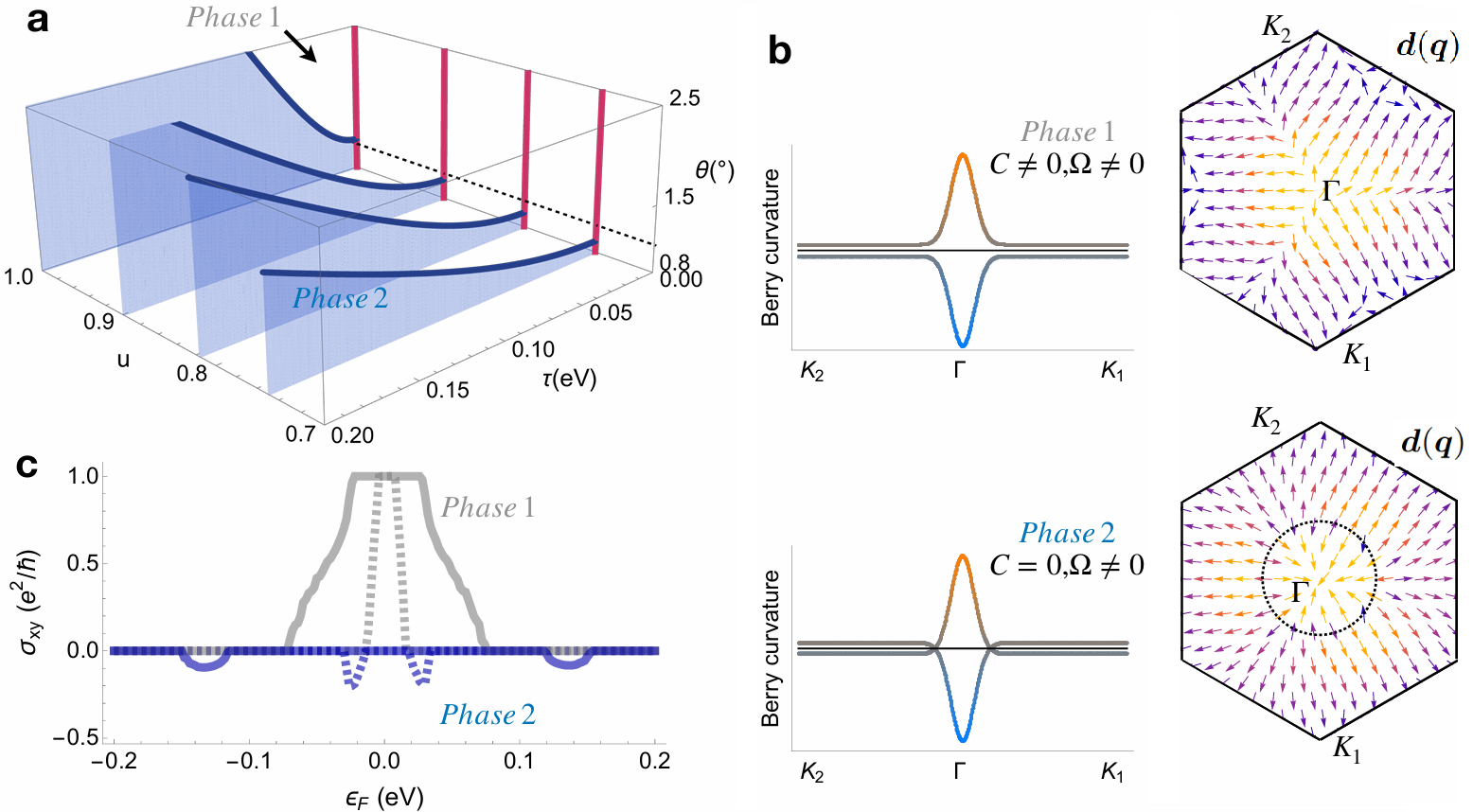}
    \caption{\textbf{(a)} 3D Phase diagram as a function of twisting angle $\theta$, time reversal symmetry breaking parameter $\tau$ and coupling $u$ with $u'=0.817$. The four 2D cuts correspond to $u=0.75,0.817,0.9,1$. The blue region is the topologically trivial phase 2 with $\mathcal{C}=0$ but $\Omega\neq 0$. The white region corresponds to the topological insulating phase 1 with $\mathcal{C}\neq 0$. \textbf{(b)} The Berry curvature along the path connecting the high-symmetry points in phase 1 (panel (b) in Fig.\ref{fig1}) is sign definite and opposite for the lowest valence and conduction flat bands. In phase 2 (panel (d) in Fig.\ref{fig1}), $\Omega$ is still peaked at the $\Gamma$ point but not anymore sign definite, leading to a vanishing total Chern number. The horizontal black line indicates $\Omega=0$ and the color map is the same as in Fig.\ref{fig1}. The vector field in the BZ is the local Berry curvature dipole $\boldsymbol{d}_n(\boldsymbol{q})$ defined in \eqref{eqdip} and computed for the lowest conduction band. The dashed black circle highlights the domain wall structure in the Berry dipole moment in phase 2. \textbf{(c)} The Hall conductivity $\sigma_{xy}$ as a function of the Fermi energy $\epsilon_F$ for phase 1 (gray color) and phase 2 (blue color). The solid gray and blue lines corresponds to panels (b) and (d) in Fig.\ref{fig1}, while the dashed ones correspond to a smaller angle, $\theta=1.1^\circ$.}
    \label{fig2}
\end{figure*}

By increasing further $\tau$ (panel (c) in Fig.\ref{fig1}), the gap at the $\Gamma$ point closes and a clear band-crossing appears in the spectrum. This gap closing and re-opening signals the onset of a second topological phase transition (TPT2) that leads to a final state (phase 2, panel (d) in Fig.\ref{fig1}) with two isolated and rather flat electronic bands with zero Chern number. This final state is topologically trivial. Nevertheless, because of the absence of time-reversal symmetry, the Berry curvature of the lowest bands with vanishing Chern number is non-zero and, on the contrary, shows a strong peak around the $\Gamma$ point (see panel (b) in Fig.\ref{fig2}). This second topological phase transition (TPT2) is similar in spirit to what recently observed by dialing the external magnetic field in TBG \cite{sanchez2024correlated}.

In panel (a) of Fig.\ref{fig2}, we show a three-dimensional phase diagram at fixed $u'=0.817$ as a function of $u,\tau,\theta$. For clarity, we only show four different 2D cuts corresponding respectively to $u=0.75,0.817,0.9,1$. The shaded colored regions indicate phase 2, while the white one represents phase 1. We observe that a larger $u$ disfavors the topological phase 1 that survives only in the region of small $\tau$ and large $\theta$. The TRSB parameter $\tau$ has an analogous effect as it pushes the system towards phase 2. On the contrary, a larger twisting angle makes phase 1 more robust and its extent in the phase diagram wider.

In panel (b) of Fig.\ref{fig2}, we plot the Berry curvature of the lowest valence and conduction bands along a q-path connecting the high-symmetry points $K_2$-$\Gamma$-$K_1$. In both cases, the Berry curvature is peaked at $\Gamma$ point. In phase 1, $\Omega$ is sign definite and has opposite sign for the lowest flat bands. On the contrary, in phase 2, $\Omega$ of the lowest bands is no longer sign definite, but it crosses the horizontal axes (black line), leading to a vanishing integral over the whole BZ, \textit{i.e.}, $\mathcal{C}=0$.

In order to characterize further the different topological features of phases 1 and 2, we use the local Berry curvature dipole density,
\begin{equation}\label{eqdip}
    \boldsymbol{d}_n(\boldsymbol{q})\equiv  \Omega_n (\boldsymbol{q}) \dfrac{\partial \epsilon_n (\boldsymbol{q})}{\partial \boldsymbol{q}} \dfrac{\partial f(\epsilon)}{\partial \epsilon},
\end{equation}
where $\epsilon_n(\boldsymbol{q})$ is the energy dispersion of the n-th band and $f$ is the Fermi-Dirac distribution function. The integral of $\boldsymbol{d}$ on the whole BZ is known to generate a nonlinear anomalous Hall coefficient even in systems with time-reversal symmetry (but broken inversion symmetry) \cite{PhysRevLett.115.216806}. In panel (b) of Fig.\ref{fig2} we show $\boldsymbol{d}$ computed on the lowest conduction band on the whole BZ for the two different phases. In phase 1, the structure of the vector field $\boldsymbol{d}$ suggests the presence of a local source of Berry dipole at the $\Gamma$ point. The situation is different in phase 2, where $\boldsymbol{d}$ shows a domain wall structure around the $\Gamma$ point (highlighted by the dashed circle) that corresponds to the location at which the Berry curvature $\Omega$ changes sign. This suggests that the topology of phase 1 can be understood from the presence of a Berry monopole at the $\Gamma$ point, while that of phase 2 presents a strong dipolar structure between opposite charges located at $\Gamma$ and $K$ points.

To connect these differences with a macroscopic measurable quantity, we consider the anomalous Hall conductivity,
\begin{equation}
    \sigma_{xy}=\dfrac{e^2}{\hbar}\int_{BZ}\dfrac{d^2 q}{(2 \pi)^2} f(\epsilon(\boldsymbol{q}))\Omega_n (\boldsymbol{q}).
\end{equation}
In panel (c) of Fig.\ref{fig2}, we show $\sigma_{xy}$ as a function of the Fermi energy for the two different phases. In phase 1, $\sigma_{xy}$ displays a plateau around $\epsilon_F=0$ where it attains the quantized value dictated by the integer Chern number $+1$, $\sigma_{xy}=e^2 \mathcal{C}/\hbar$. The width of the plateau is controlled by the flatness of the lowest topological bands and it decreases moving towards the magic angle (dashed gray line). On the other hand, $\sigma_{xy}$ is not quantized in phase 2 but, despite $\mathcal{C}=0$, it is nonzero and negative when the Fermi energy crosses the lowest flat bands. Again, if these bands have a more flat dispersion, as expected for twisting angles closer to the magic one, $\sigma_{xy}$ acquires a larger value and a sharper form. We emphasize that, for the phase at $\tau=0$, $\sigma_{xy}$ vanishes for any $\epsilon_F$ since the Berry curvature is identically zero as a consequence of time reversal symmetry. The different behavior of $\sigma_{xy}$ in the three phases suggests that they can be distinguished in the lab and the corresponding properties can be connected to the topological features of their electronic spectrum.

In summary, we discovered a time-reversal symmetry breaking induced topological phase transition in TBG separating a topological insulating phase and a topological trivial phase in which the spectrum exhibits a pair of isolated bands with zero Chern number but nonzero local Berry curvature. We have shown how to identify this new phase in the lab by measuring a non-zero, and not quantized, anomalous Hall conductivity upon tuning the Fermi energy across the lowest isolated bands and related this property to the topological Berry curvature dipole.

We provided a clear demonstration of how TRSB can help engineering and controlling the topological properties of TBG \cite{Adak2024} leading to novel electronic phases. We expect that our findings might be generally applicable to several 2D moir\'e systems and that this novel topological phase transition could be experimentally detected in the near future.

\section*{Acknowledgments}
The authors thank Jinhua Gao and Giandomenico Palumbo for useful discussions. CJ and MB acknowledge the support of the Shanghai Municipal Science and Technology Major Project (Grant No.2019SHZDZX01). MB acknowledges the support of the sponsorship from the Yangyang Development Fund.
QDJ sponsored by National Natural Science Foundation of China (NSFC) under Grant No.12374332, Jiaoda2030 Program Grant No.WH510363001, TDLI starting up grant, and Innovation Program for Quantum Science and Technology Grant No.2021ZD0301900.

\end{document}